\documentclass[]{spie}  

\usepackage{amsmath,amsfonts,amssymb}
\usepackage{graphicx}
\usepackage[colorlinks=true, allcolors=blue]{hyperref}

\usepackage[dvipsnames]{xcolor}

\title{Breaking the brightness barrier: JWST/NIRCam DHS spectroscopy for high-precision time-series observations}

\author[a]{Achrène Dyrek}
\author[a]{John Stansberry}
\author[a]{Louis E. Bergeron}
\author[b]{Everett Schlawin}
\author[a]{Nestor Espinoza}
\author[a]{Brian Brooks}
\author[a]{Mario Gennaro}
\author[a]{Martha L. Boyer}
\author[a]{Russell Ryan}
\author[a]{Bryan Hilbert}
\author[a]{Munazza K. Alam}
\author[a]{Aarynn L. Carter}
\author[a]{Norbert Pirzkal}
\author[a]{Julien H. Girard}
\author[a]{Anton M. Koekemoer}
\affil[a]{Space Telescope Science Institute, 3700 San Martin Drive, Baltimore, MD 21218}
\affil[b]{Astrophysics \& Space Center, Schmidt Sciences, New York, NY 10011, USA}

\authorinfo{Further author information: (Send correspondence to A. Dyrek)\\A. Dyrek: E-mail: adyrek@stsci.edu}

\begin{document}
\pagestyle{empty}
\maketitle

\begin{abstract}
Many of the most scientifically compelling exoplanets orbit bright nearby stars that exceed the brightness limits of existing JWST spectroscopic observing modes. To address this limitation, a new NIRCam Short Wavelength Grism Time-Series mode has been developed by combining the Dispersed Hartmann Sensor (DHS) with a new on-board multistripe detector readout capability. The DHS disperses the incoming light through multiple pupil sub-apertures, reducing the incident flux and providing slitless spectroscopy between approximately 1.0 and 2.3~$\mu$m. The multistripe readout mode further increases the accessible brightness range ($K\sim2.5-5.7$ mag, depending on the spectral type of the object), with the previous limit being $K\sim$5.7 mag at 1.5$\mu$m, by reading only the detector regions containing the DHS spectra, reducing the detector frame time with the standard RAPID readout mode from 10.74~s to 1.36~s over even less when using smaller substripe sizes. Together with the simultaneous long-wavelength grism observations, the new mode provides spectroscopic continuous coverage from approximately 1.0 to 5.0~$\mu$m for targets as bright as $K\sim2.5$ mag. We present the first results of on-orbit commissioning of this new observing mode with the final stage consisting of observations of a full transit of the exoplanet WASP-18b, which allowed us to demonstrate the feasibility of using NIRCam DHS for high-precision time-series observations. 
The commissioning presented in this paper marks the first deployment of the multistripe detector readout on JWST. Beyond NIRCam DHS, this new capability will be extended to other spectroscopic modes, including NIRISS SOSS and NIRSpec PRISM. Bright nearby stars host many of the highest-priority targets for exoplanet atmospheric characterization, making multistripe an important new capability for maximizing the scientific return of JWST.
\end{abstract}


\section{SCIENTIFIC MOTIVATION}
\label{sec:intro}  

The characterization of planetary atmospheres through spectroscopy is one of the primary scientific goals of the James Webb Space Telescope (JWST). Observations spanning the near- and mid-infrared enable the detection of molecular absorption features such as H$_2$O, CO$_2$, CO, CH$_4$, NH$_3$, and SO$_2$, providing direct constraints on atmospheric composition, chemistry, clouds, and thermal structure. Since the beginning of science operations, JWST has demonstrated unprecedented spectroscopic precision across all four science instruments, producing detailed transmission and emission spectra of exoplanet atmospheres that reveal a remarkable diversity of atmospheric properties \cite{espinoza_highlights_2025}.

Exoplanet transit and eclipse measurements require continuous observation of a target over several hours while maintaining exceptional photometric and spectroscopic stability. During such time-series observations, spectra are acquired continuously before, during, and after the planetary transit or eclipse, allowing the small decrease in stellar flux produced by the planetary atmosphere to be measured. Spectroscopic observations allow characterization of the composition of the exoplanet atmosphere because the radial extent of the atmosphere, and therefore transit depth, depends on absorption bands of the molecular constituents. Broader wavelength coverage provides inherently stronger constraints on the composition and structure of the atmosphere. Achieving photon-noise-limited precision over these long observations is therefore one of the principal requirements of JWST's dedicated time-series observing modes \cite{greene_characterizing_2016}.

Many of the most scientifically compelling exoplanet systems orbit some of the brightest stars in the sky. Nearby systems such as 55 Cancri, HD 219134, and $\nu^2$ Lupi host terrestrial or sub-Neptune planets that are among the highest-priority targets for atmospheric characterization, yet their host stars exceed the brightness limits of many standard JWST observing modes. For these targets, the primary limitation comes from the way the JWST detectors are read. Rather than taking a single measurement at the end of an integration, the detectors are read repeatedly during the exposure using a technique called \textit{up-the-ramp} sampling. The time between two successive reads determines how much signal a pixel can accumulate before it is measured again. For very bright stars, pixels fill with photoelectrons faster than they can be read, causing them to saturate before the end of the exposure and preventing high-precision spectroscopic measurements. Extending JWST observations toward brighter stars therefore represents an important step toward characterizing nearby planetary systems that are inaccessible using existing observing modes.

Several approaches can be used to increase the brightness limits of infrared detectors, including reducing the collected flux, dispersing the incoming light over more detector pixels, and decreasing the detector frame time through faster readout schemes (which can include the use of subarrays). NIRCam naturally provides the first of these through its Dispersed Hartmann Sensor (DHS), originally designed as a wavefront sensing element \cite{rigby_science_2023} and now to be used for spectroscopy. To fully exploit the DHS for time-series spectroscopy, a new detector readout pattern, referred to as the multistripe readout mode, was developed and implemented on board JWST. Instead of reading out the entire detector every frame, the multistripe mode rapidly reads only the narrow detector regions containing the DHS spectra while skipping unused detector rows. This technique of reading one smaller region of the detector (which we call a subarray) to reduce the frame time has been extensively used with JWST. The multistripe mode brings a novel approach by reading not only one but several subarrays of the detector in one frame (see Sect.~\ref{sec:multistripe}). This reduces the detector frame time by nearly an order of magnitude relative to the standard full-frame RAPID readout, significantly increasing the brightness limit accessible with NIRCam spectroscopy. At 1.5 $\mu$m, our previous brightness limit was $K\sim$ 5.7 mag for a solar-type star, considering two reads of the 160$\times$160 subarray. 

The combination of DHS spectroscopy with the new multistripe readout mode establishes a new NIRCam observing mode: the Short Wavelength Grism Time-Series mode. For time-series spectroscopy, NIRCam has traditionally relied on the long-wavelength (LW) grisms, which provide slitless spectroscopy over the wavelength range of approximately 2.4--5.0~$\mu$m, paired with short-wavelength (SW) photometry. The new observing mode presented in this paper extends NIRCam spectroscopy into the SW channel, providing simultaneous spectroscopic coverage between approximately 1.0 and 2.3~$\mu$m while preserving the standard LW grism observations\cite{schlawin_two_2017}.

This paper presents the on-orbit commissioning of the new NIRCam Short Wavelength Grism Time-Series observing mode. The commissioning campaign consisted of three complementary observing programs. PID 4453 demonstrated DHS spectroscopy using the existing full-frame detector readout and established the detector geometry. Following the successful implementation of the multistripe readout on board JWST, PID 9215 obtained observations of wavelength and flux calibration standard targets covering both LW and SW. Finally, PID 9243 acquired the first high-precision transit time-series using the new observing mode through observations of the hot Jupiter WASP-18b, demonstrating its stability and readiness for community science observations.

This paper is organized as follows. Section~2 introduces the NIRCam DHS spectroscopic configuration and defines the field points, filters, and detector geometry used by the new observing mode. Section~3 describes the multistripe detector readout and its implementation on board JWST. Section~4 summarizes the commissioning campaign and presents the current calibration status. Section~5 presents the first on-sky time-series demonstration with observations of the exoplanet WASP-18b, and Section~6 summarizes the capabilities of the new observing mode and its availability for future JWST programs.

\section{ENABLING NIRCAM DHS SPECTROSCOPY}
\label{sec:dhs_spectroscopy}

The DHS was originally developed as part of the NIRCam wavefront sensing system and played an important role during JWST commissioning, where it was used to measure and achieve coarse phasing of the segments of the primary mirror. The DHS consists of ten separate grisms mounted within small sub-apertures in NIRCam's pupil wheel. Each sub-aperture samples only two primary-mirror segments and therefore transmits only a few percent of the total JWST collecting area, reducing the total amount of light reaching the detectors. Further, each DHS grism produces a separate spectrum, displaced cross-dispersion on the detectors. Finally, the spectral dispersion ($\sim$300 for the first order) also reduces the flux incident on the detectors. In sum, these effects significantly enhance the bright limit in NIRCam's SW channel. A drawback is that the DHS produces 10 spectra of every source in the field of view (defined by the extent of the pickoff-mirror), so its use is best restricted to observations of well-isolated bright targets.

\begin{figure}[htbp]
\centering
\includegraphics[width=0.7\textwidth]{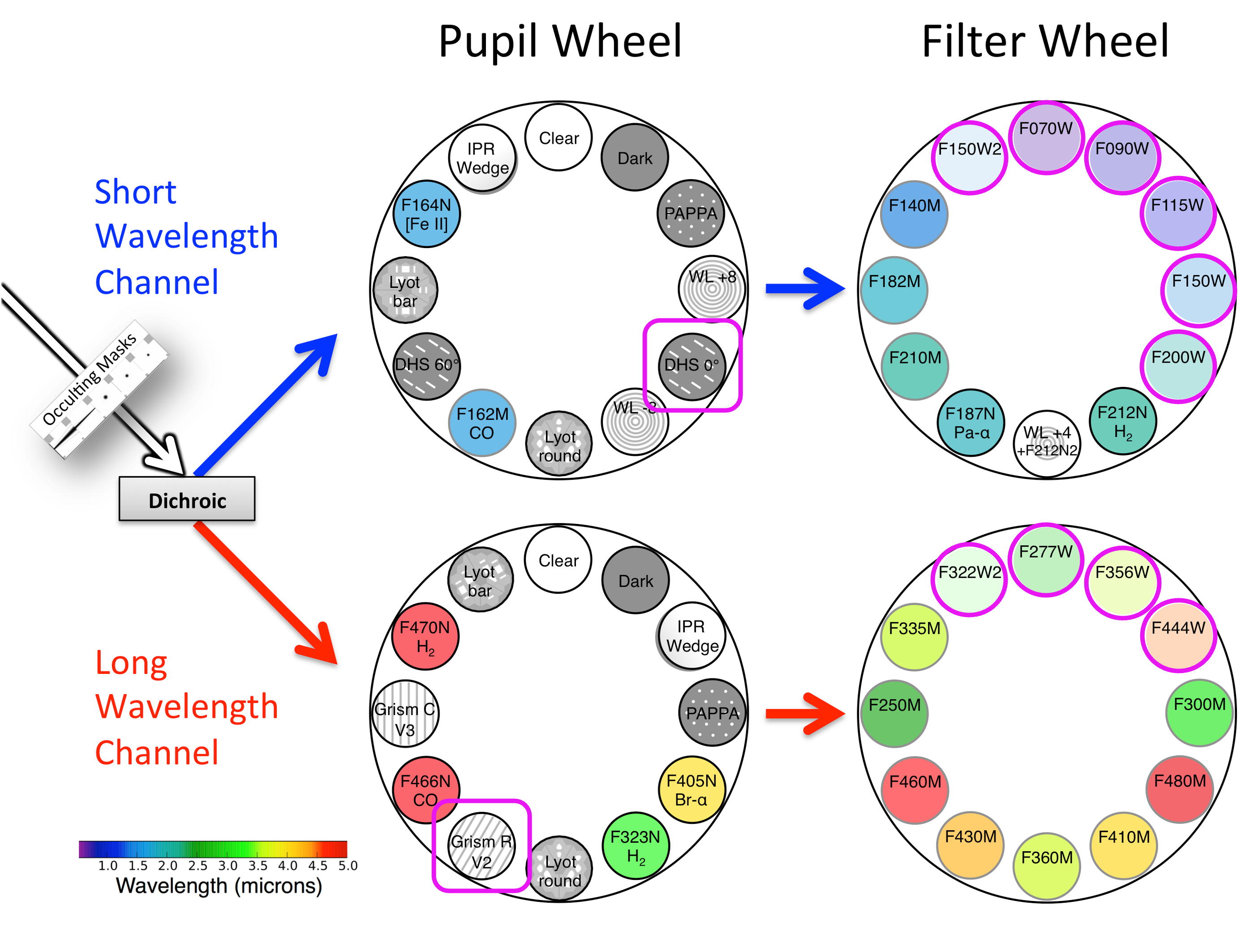}
\caption{NIRCam short- and long-wavelength pupil and filter wheels. The optical elements used for NIRCam DHS observations are highlighted in pink, including the short-wavelength DHS element, the short-wavelength blocking filters, the long-wavelength grisms, and the corresponding long-wavelength filters. }
\label{fig:dhs_optics}
\end{figure}

Figure~\ref{fig:dhs_optics} illustrates the optical elements used for DHS observations, including the SW and LW channels. The DHS is located in the NIRCam short-wavelength pupil wheel. When selected, the incoming beam passes through the DHS element before reaching one of the six available short-wavelength blocking filters (F070W, F090W, F115W, F150W, F150W2, or F200W) in the filter wheel. The DHS consists of ten grating-equipped sub-apertures that each intercept a small fraction of the telescope pupil, producing multiple dispersed spectra on the short-wavelength detectors of NIRCam Module~A (NRCA1--NRCA4), where each illuminated DHS sub-aperture produces an independent spectrum. Two DHS pupil elements exist, denoted DHS 0$^{\circ}$ and DHS 60$^{\circ}$, corresponding to the orientation of the dispersion direction on the detector. The new spectroscopic mode uses the DHS 0$^{\circ}$ element, which disperses the spectra approximately along the detector rows.

Although the DHS contains ten sub-apertures, only eight are used for science observations as two of the sub-apertures (sub-apertures 1 and 6) fall within the short-wavelength detector gap. During the development of the spectroscopic mode, the rotation of the DHS was re-optimized such that the spectral traces would be parallel (to the extent possible) with rows of pixels on the detectors. This allowed for narrower subarray substripes and thereby shorter frame times (see Sect.~\ref{sec:multistripe}). The 8 science DHS sub-apertures do not all transmit the same amount of light, as their projected areas on the JWST primary mirror differ slightly. Their fractional collecting areas range from 2.67\% to 3.94\%. Together, the 8 science sub-apertures collect approximately 25.8\% of the telescope pupil, providing a substantial reduction in incident flux compared to the full aperture while maintaining high signal-to-noise spectroscopy for very bright sources.

\begin{figure}[htbp]
\centering
\includegraphics[width=0.7\textwidth]{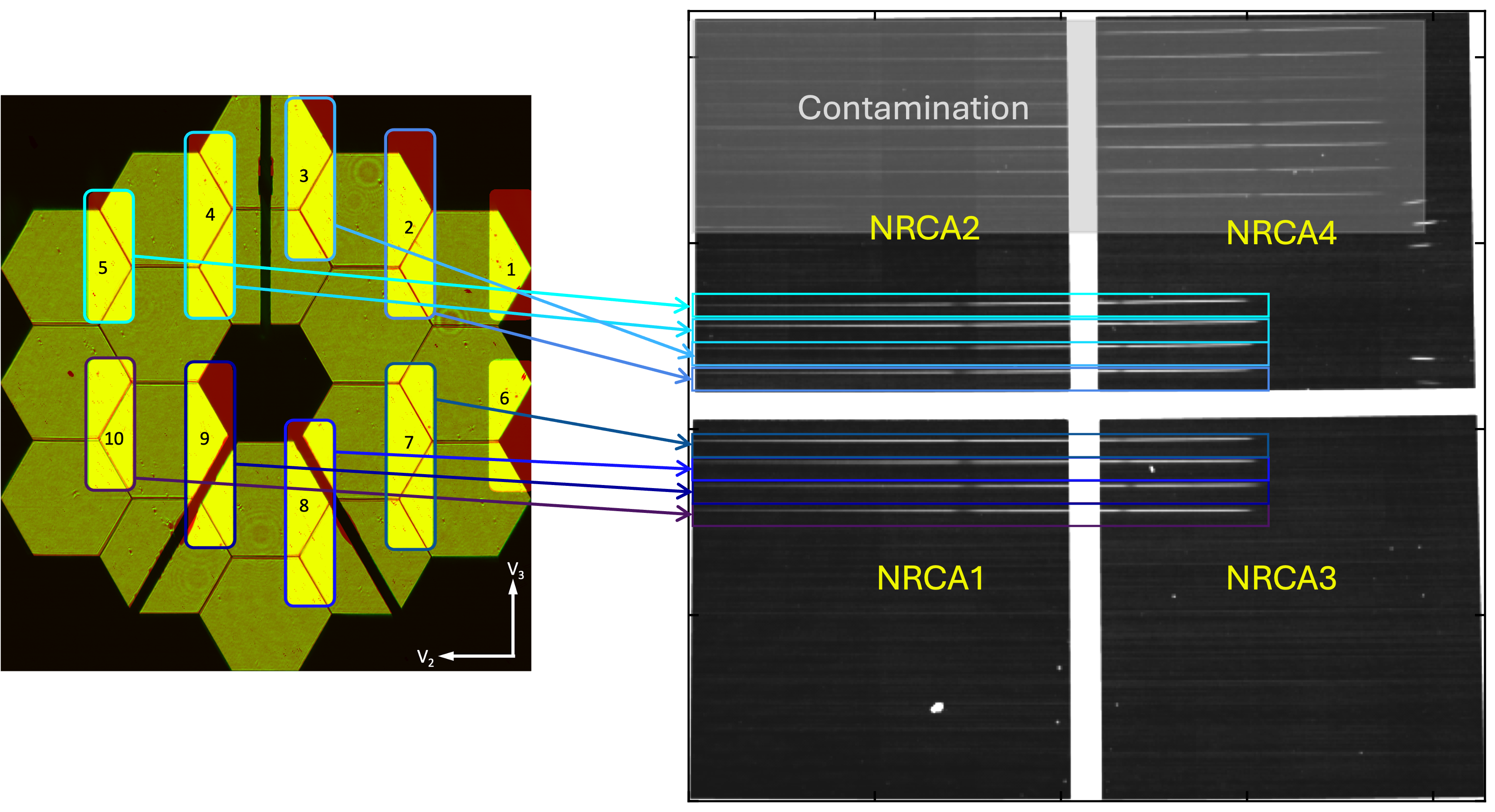}
\caption{The NIRCam DHS spectroscopic configuration and detector footprint. \textit{Left:} The 10 DHS science sub-apertures projected onto the JWST primary mirror, with sub-apertures 1 and 6 projecting onto the detectors' horizontal gap, using filter F150W2 and field-point F444W. \textit{Right:} The corresponding first-order spectra on the four NIRCam Module A short-wavelength detectors. The shaded region indicates an example of spectral contamination from a nearby astrophysical source. Bright blocks on detectors NRCA1 and NRCA3 are bad pixels.}
\label{fig:mirror_spectra}
\end{figure}

Similarly to the imaging time-series capability on NIRCam, DHS spectroscopy uses both the short- and long-wavelength channels simultaneously. The short-wavelength spectra are recorded on the four short-wavelength detectors, while the corresponding long-wavelength spectrum is acquired simultaneously using the long-wavelength grism, on one detector. Consequently, every DHS observation naturally provides spectroscopic information in both channels without requiring additional exposures. Four spectroscopic configurations, referred to as field points, are available. These correspond to the two long-wavelength grism settings, F322W2, F444W, F277W and F356W, and determine the location of the target within the NIRCam field of view and the resulting positions of the spectra on the short-wavelength detectors.  Each field point therefore produces a different detector footprint and wavelength coverage. Although these 4 long-wavelength filters are available, only F322W2 and F444W are part of our current commissioning plan. For a given field point, each of the 8 DHS science sub-apertures produces an independent dispersed spectrum on the detectors (Fig.~\ref{fig:mirror_spectra}) of the same target. The individual DHS grisms have small tilts parallel to the dispersion direction, so each spectrum has a unique wavelength solution.  

As for all slitless spectroscopic modes, the placement of the dispersed spectra on the detector introduces several observing constraints. First, for the F150W2 filter and F322W2 field-point combination, second-order spectral contamination may overlap portions of the first-order spectra (see Sect.~\ref{sec:results}). The affected wavelength regions are identified during the wavelength calibration process and should be considered during scientific analysis. A second consideration comes from the wide-field nature of the DHS observations. Since no slit is used, dispersed spectra from neighboring field sources may overlap the target spectra depending on the telescope orientation. Consequently, the telescope position angle (PA) should be selected to minimize contamination from nearby sources whenever possible. Appropriate PA constraints depend on the surrounding field and the selected field point, as the orientation of the dispersed spectra differs between the two DHS configurations. The \texttt{ExoCTK} tool, which is an open-source data analysis package on exoplanetary atmospheric characterization and time-series observations planning, includes a contamination tool to help assess contamination from nearby sources\footnote{\url{https://exoctk.stsci.edu/contam_visibility}}.

\section{THE MULTISTRIPE READOUT MODE}
\label{sec:multistripe}

Although the DHS significantly reduces the amount of light reaching the short-wavelength detectors, the detector readout speed remains the dominant limitation for the brightest targets. In the standard NIRCam full-frame RAPID readout mode, a complete frame requires 10.74~s to read. During this time, photoelectrons continue to accumulate in every pixel. For bright stars, the detector can therefore reach saturation before the next detector read, preventing accurate measurements and limiting the achievable brightness of observations.

Reducing the frame time is therefore crucial to increase the brightness limit. Since the DHS spectra fall on only a small fraction of the detector area (Fig.~\ref{fig:mirror_spectra}), reading the full $2048\times2048$ detector every frame is unnecessary. In principle, all 4 DHS spectra on each detector could be read altogether as a unique big subarray of about 2048$\times$512 pixels, with a frame time of 2.7 s. To further decrease the frame time, only detector rows containing the spectra could be read, while the remaining rows could be skipped. This concept motivated the development of a new family of detector readout operations collectively referred to as multistripe. In fact, the observing mode described in this paper uses the multistripe substripe implementation, although throughout this paper it is referred to simply as multistripe.

The multistripe readout modifies the standard detector clocking sequence by alternating between detector regions that are read and regions that are skipped. After reading a stripe of detector reference pixels, one or more additional stripes of pixels containing DHS spectra are read while rapidly skipping over detector rows that contain no useful signal. This sequence is repeated for each DHS stripe until all spectra have been sampled. If multiple reads are specified for an integration, the pattern repeats until the integration is complete. 

\begin{figure}[htbp]
\centering
\includegraphics[width=0.7\textwidth]{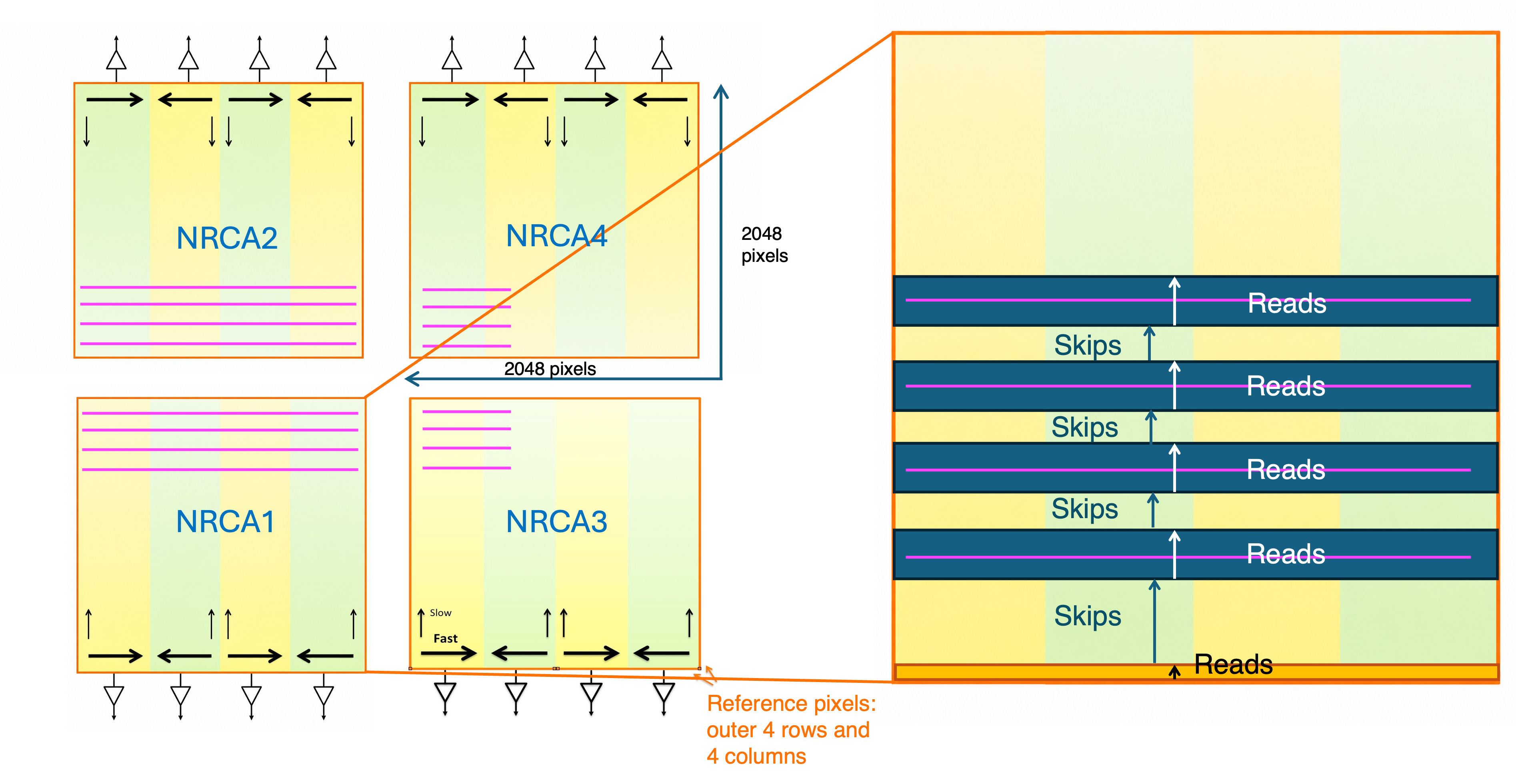}
\caption{Diagram of the multistripe readout mode with schematic stripes mimicking the output for the F150W2 filter and F444W field-point combination. \textit{Left:} Layout of the DHS spectra on the four NIRCam Module A short-wavelength detectors. Only the detector rows containing the spectra are selected for readout. \textit{Right:} Schematic of the multistripe readout sequence on one detector, in which the detector alternates between reading the selected stripes and skipping the intervening rows before returning to the reference pixels at the bottom of the detector.}
\label{fig:multistripe+full}
\end{figure}

Figure~\ref{fig:multistripe+full} illustrates the principle of the multistripe readout. Only the detector rows containing the DHS spectra are digitized, while the intervening detector rows are traversed without recording science pixels. Implementing this new readout required modifications to the detector readout firmware rather than simply defining a new subarray. The subtlety of such a new mode resides in the fact that the clocking system must go back to reference pixels and read them before starting to read a new science stripe. The detector Application Specific Integrated Circuit (ASIC) microcode was updated to support alternating read and skip operations while maintaining the detector timing, reference pixel sampling, and synchronization with the long-wavelength channel. Following extensive ground testing, the new microcode was uploaded and successfully commissioned on board JWST, establishing a new family of detector readout modes for NIRCam time-series spectroscopy. This way, the standard configuration reads all eight DHS spectra and provides a frame time of 1.36~s. Shorter frame times are available if fewer substripes are specified (thus collecting signal from fewer DHS spectra), the shortest being 0.22 s if only the 2 brightest spectra are read out.

In addition to the standard multistripe configuration, several smaller subarrays are available to further increase the brightness limit by reducing the number of spectra and detector rows that are read. The standard \texttt{SUB260S4\_8-SPECTRA} configuration reads all eight DHS spectra, with each stripe spanning 65 detector rows. For brighter targets, the \texttt{SUB164S4\_8-SPECTRA} configuration records only the four central DHS spectra using 41 rows per stripe. The \texttt{SUB82S2\_4-SPECTRA} and \texttt{SUB41S1\_2-SPECTRA} configurations further reduce the number of spectra to the two central DHS spectra, each occupying 41 detector rows. Reading fewer stripes decreases the detector frame time and extends the brightness range accessible with NIRCam time-series spectroscopy. The smaller subarrays retain the central DHS spectra, which provide the highest throughput and most complete wavelength coverage, while omitting the outer spectra to reduce the detector area that must be read.

While the smaller subarrays provide shorter frame times, they also reduce the amount of detector area surrounding the spectra. This can limit the number of pixels available for estimating and subtracting the background, which may be important for some time-series observations. Consequently, observers should balance the need for shorter frame times against the availability of sufficient background pixels when selecting a subarray. Another important consideration is that saturation typically occurs first in the long-wavelength channel, making it the limiting factor when selecting the observing configuration and readout pattern.

In DHS mode, 5 detector arrays are being read simultaneously (4 SW detectors and 1 LW detector). This substantially increases the data volume and may exceed the available solid-state recorder capacity for long integrations. To mitigate this limitation, additional detector readout patterns (DHS3--DHS7) were developed. These patterns periodically skip complete detector frames while preserving the rapid detector sampling required for bright targets, thereby reducing the total data volume without modifying the detector integration scheme.

\section{COMMISSIONING CAMPAIGN}
\label{sec:results} 

The first on-sky commissioning observations of the DHS spectroscopic mode were obtained as part of program PID 4453 (PI: J. Stansberry) before the implementation of the multistripe readout mode. These observations used the standard full-frame detector readout and had three primary objectives. First, they verified that the dispersed spectra produced by the 8 DHS science sub-apertures were located at their predicted detector positions. Second, they validated the detector geometry required for the future multistripe readout by measuring the locations and spatial extent of each spectrum. Finally, the observations tried the new DHS observing configurations and detector subarrays to verify their functionality on orbit.

\begin{figure}[htbp]
\centering
\includegraphics[width=0.8\textwidth]{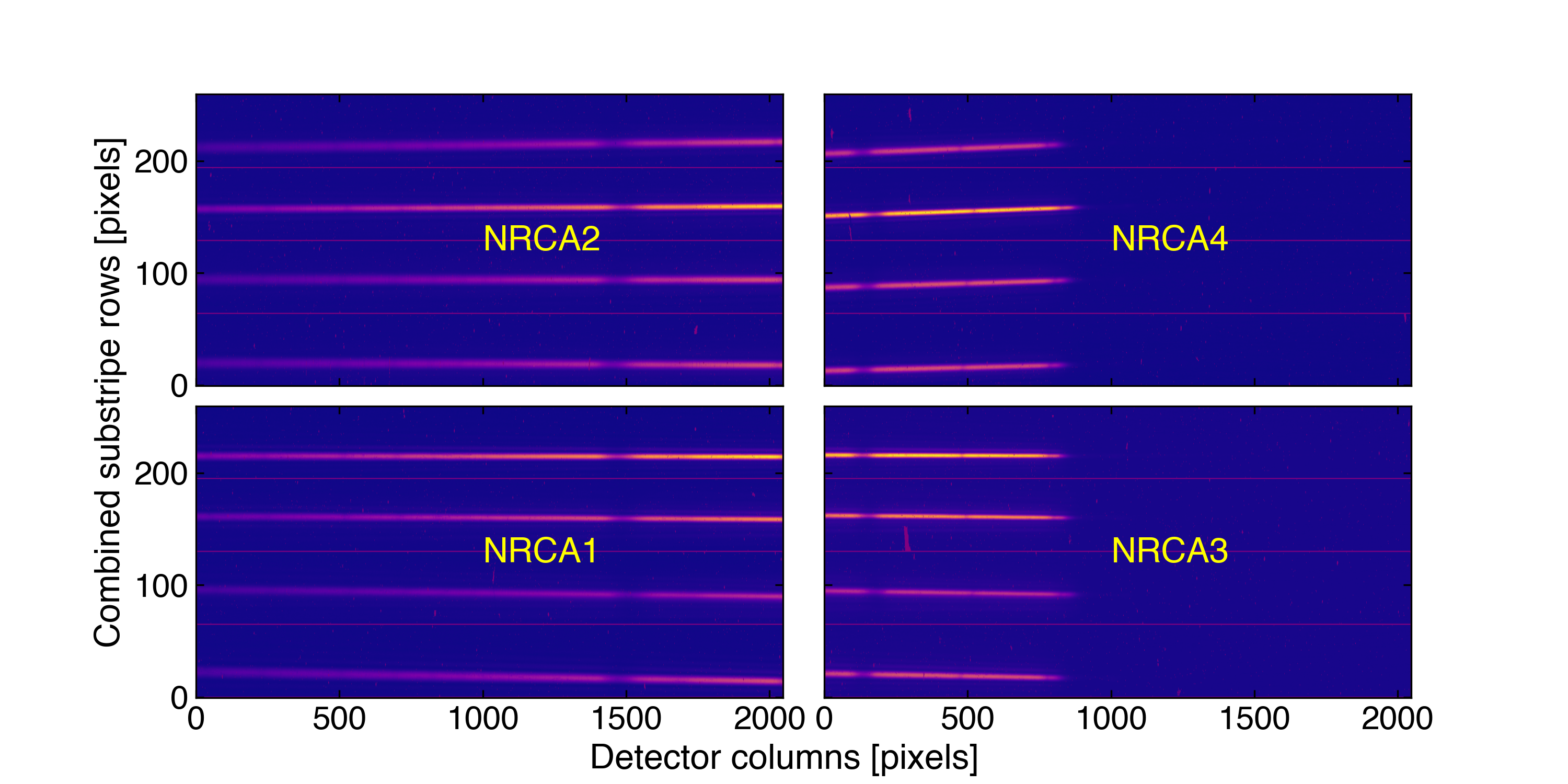}
\caption{Spectral traces on all four NIRCam detectors of calibrator star $\eta^1$ Doradus for the F150W2 filter and F444W field point, using the multistripe readout mode with the \texttt{SUB260S4\_8-SPECTRA} configuration. The vertical offsets of the traces between the left and right sides are due to the physical alignment of the 4 detectors in the focal plane.}
\label{fig:2D_spectra}
\end{figure}

Following the successful commissioning of the multistripe readout in March 2026, program PID 9215 (PI: J. Stansberry) was designed to establish the wavelength and flux calibration of the new observing mode. Observations were obtained using the standard  \texttt{SUB260S4\_8-SPECTRA} configuration in the RAPID readout mode for the two F444W and F322W2 DHS field points and all six short-wavelength blocking filters. Two spectrophotometric standard stars (LAWD-52 and 2MASS-J17430448+6655015) as well as the Large Magellanic Cloud planetary nebula IRAS-05248-7007 were observed to provide absolute flux calibration over the full wavelength range, while the A0V star $\eta^1$ Doradus served as the primary wavelength calibrator because of its numerous hydrogen absorption features. Figure~\ref{fig:2D_spectra} shows the resulting spectral traces on the four detectors for the F444W field point and F150W2 filter. 

A dedicated calibration pipeline has been developed to process the new DHS observations. The standard JWST Stage-1 calibration pipeline first produces the \texttt{rateints} data products, after which we designed a dedicated extraction algorithm to identify the individual DHS traces, determine their positions on the detector, and extract one-dimensional spectra for each detector, substripe, and integration. The extracted spectra were subsequently used to derive wavelength solutions for both field points F444W and F322W2 and all six filters. Trace products were stored in detector coordinates, while the wavelength solutions were derived in the DHS science coordinate system to remain consistent with the NIRCam calibration conventions. The calibration procedure and extraction algorithm are described in detail in the accompanying technical report to be submitted soon.

Because the DHS traces may be incomplete, partially contaminated, or truncated by the detector boundaries, the extraction algorithm was designed to identify the spectral traces directly from the detector images without relying on predetermined trace locations. Individual traces were detected independently for every integration, after which a polynomial model describing the trace geometry was derived and used to perform subpixel-weighted spectral extraction. In contrast to conventional slitless spectroscopy modes that rely on a direct image to determine the source position, the DHS extraction identifies the spectral traces directly in the dispersed images. The trace locations are measured independently for each detector and substripe and subsequently used for spectral extraction and wavelength calibration. Because the target is placed at a predefined DHS field point, the observed trace positions were found to be highly repeatable throughout the commissioning observations.

\begin{figure}[htbp]
\centering
\includegraphics[width=0.45\textwidth]{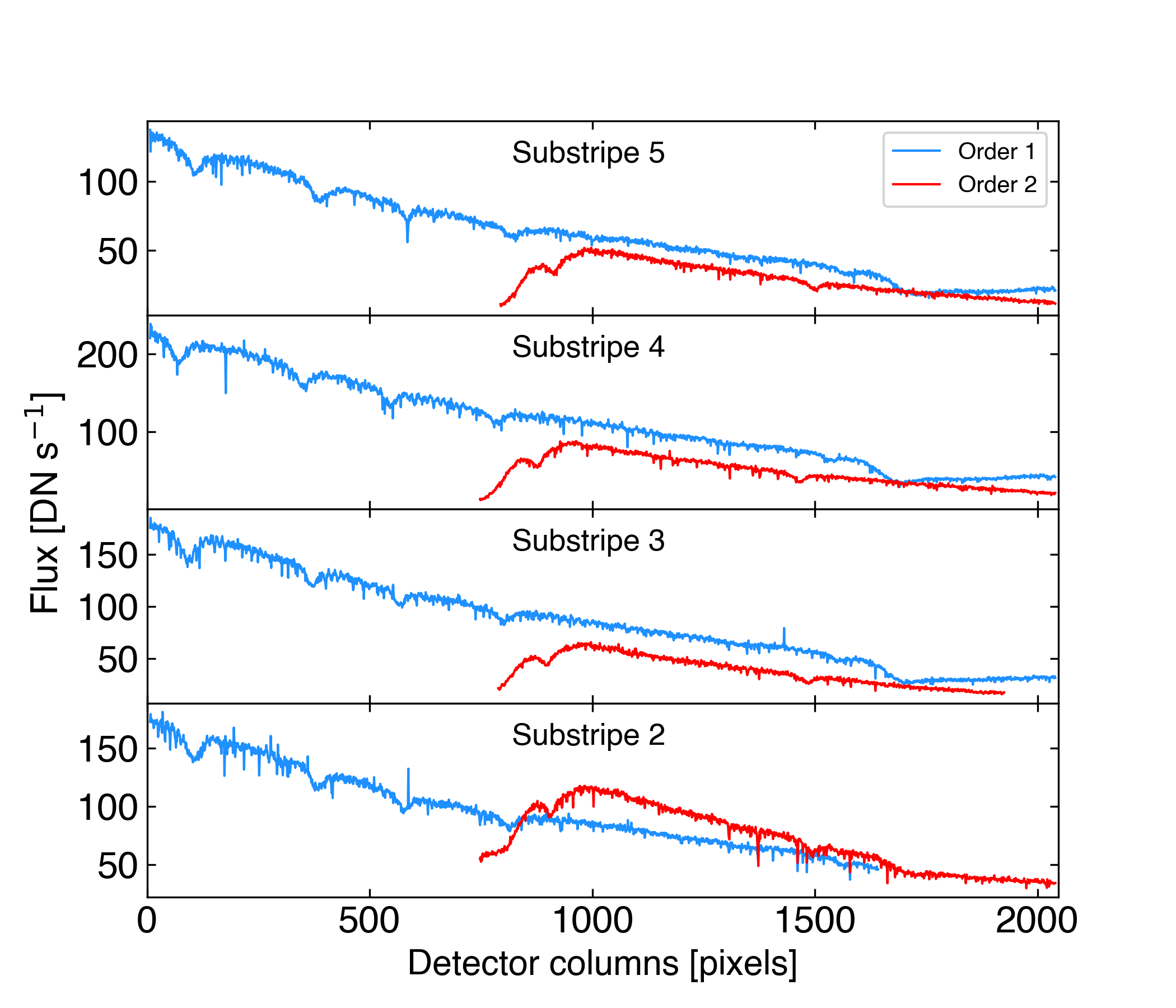}
\includegraphics[width=0.45\textwidth]{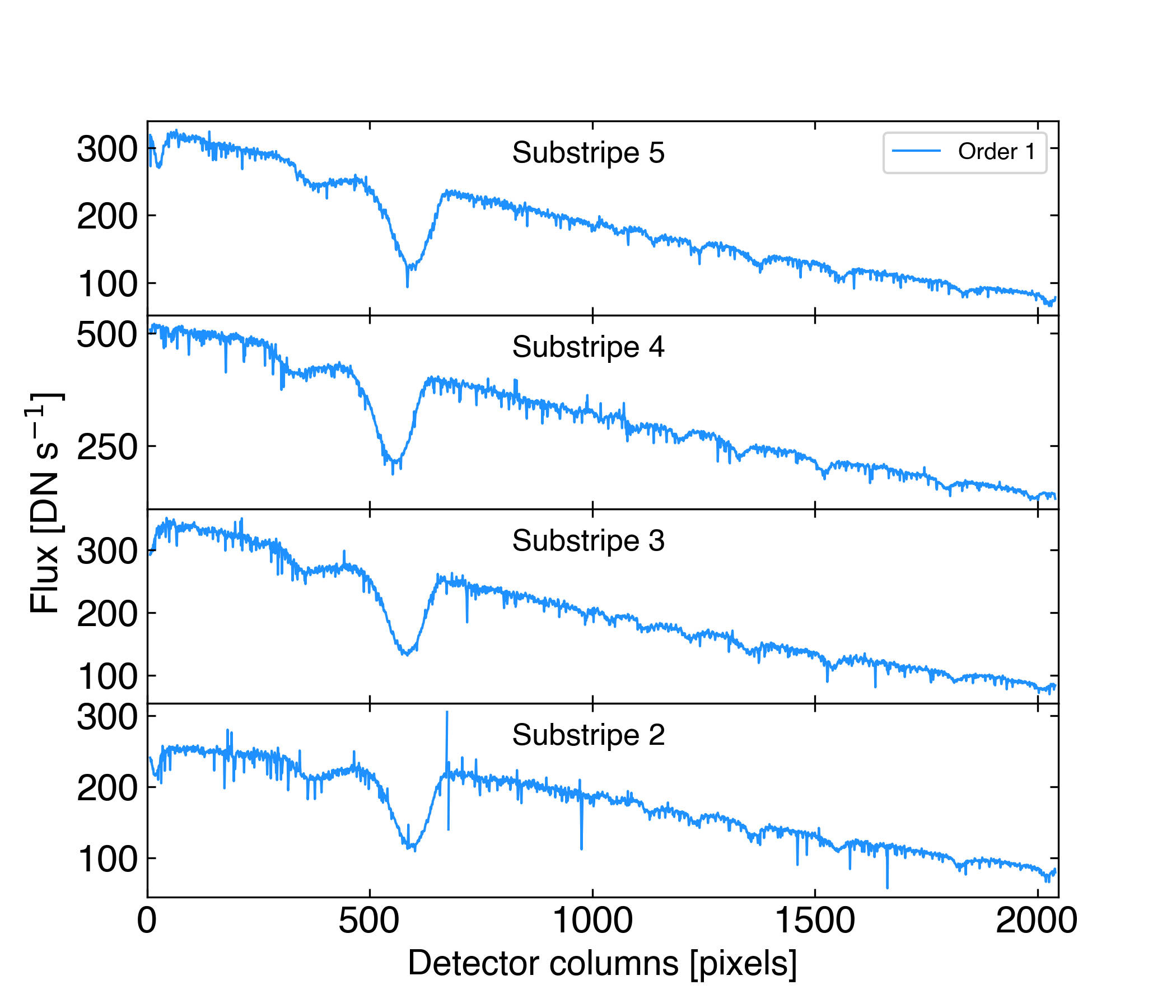}
\caption{Representative extracted one-dimensional spectra of calibrator star $\eta^1$ Doradus for the F150W2 filter on detector NRCA2. \textit{Left:} F322W2 field point. The first-order spectrum (blue) is contaminated by second-order light (red). \textit{Right:} F444W field point. The first-order spectra are free of second-order contamination over the full wavelength range.}
\label{fig:1D_spectra}
\end{figure}

Figure~\ref{fig:1D_spectra} illustrates representative extracted spectra for detector NRCA2 and the two field points F444W and F322W2. The F444W configuration provides largely uncontaminated first-order spectra, whereas the F322W2 field point exhibits overlap with second-order light at the longest wavelengths for several DHS spectra. In the current calibration, the affected wavelength region is excluded from the recommended science wavelength range.

The detector geometry of the DHS mode has now been validated for both field points and all six short-wavelength filters. The wavelength calibration has been established for the \texttt{SUB260S4\_8-SPECTRA} configuration and is currently being propagated to the remaining DHS spectra and observing configurations. The modified calibration pipeline successfully processes the new multistripe observations, including trace identification and spectral extraction from the individual substripes. Flux calibration is currently being finalized using observations of the spectrophotometric calibration standards. At the time of writing, the F150W2+F444W configuration has completed commissioning and is available for General Observer (GO) observations, and the F150W2+F322W2 configuration is expected to follow.

\section{FIRST TIME-SERIES DEMONSTRATION}
\label{sec:dhs_timeseries}

To demonstrate the performance of the new DHS observing mode for exoplanet time-series observations, a full transit observation was acquired as part of program PID 9243 (PI: A. Dyrek). The commissioning target was selected to maximize the ability to assess the instrumental performance of the DHS mode rather than to address a specific scientific question. In fact, the target was chosen for its absence of atmospheric features. This way, all detected features can be attributed to instrumental systematics. Three performance metrics were identified prior to the observations: (1) the photometric precision of the extracted light curves, (2) the uncertainty on the measured transit depth, which determines the achievable precision of transmission spectra, and (3) the level of correlated (red) noise remaining in the light-curve residuals. These metrics were used to guide the selection of the commissioning target and the observing strategy.

A sample of known transiting exoplanets observable during the 2026 commissioning window was evaluated using scheduling constraints, detector saturation limits, transit duration, transit depth, stellar activity, and contamination from nearby sources. WASP-18b emerged as the preferred target because it satisfies all operational constraints while providing a relatively deep transit around a bright host star (K = 8.1 mag) with a short observing duration ($<$ 6 hours). The observations were obtained in June 2026 using the standard DHS time-series \texttt{SUB260S4\_8-SPECTRA} configuration with the F150W2 short-wavelength filter, the multistripe DHS3 readout pattern and the F444W long-wavelength field point. WASP-18b was observed from June 06 2026 16:32:37 UTC to June 06 2026 20:37:59 UTC with a total duration of 4.08 hours and a transit duration of 2.21 hours. The data is split into 3 uncalibrated files per detector with a total of 371 integrations of 10 frames each. This DHS TSO observation marks the first science configuration commissioned for community use.  Contemporaneous observations with the Hubble Space Telescope (HST) Wide Field Camera 3 (WFC3) will also be obtained as part of program PID 18234 (PI: A. Carter) to provide an independent comparison of the measured transit depth and instrumental systematics.

\begin{figure}[htbp]
\centering
\includegraphics[width=0.6\textwidth]{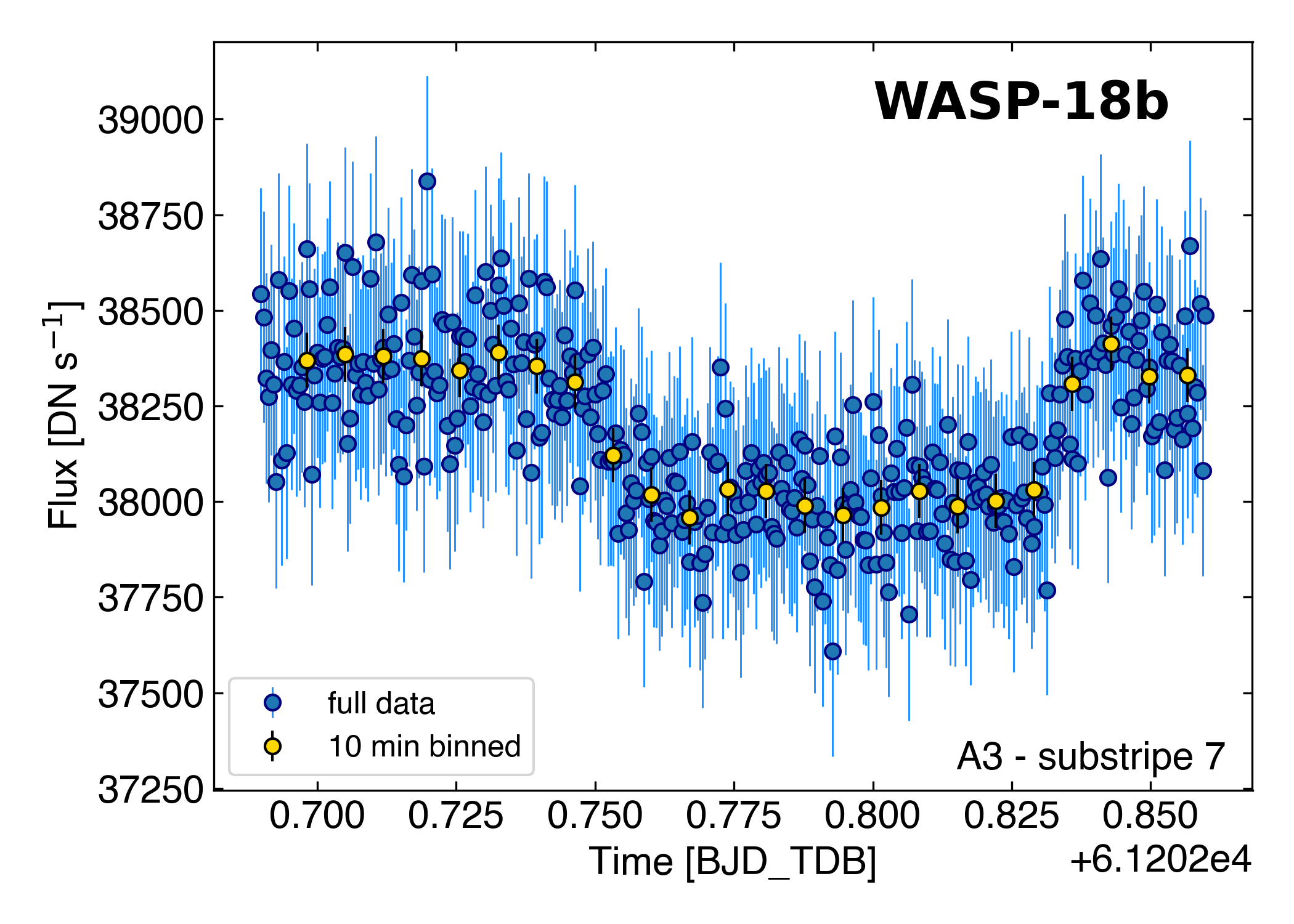}
\caption{White-light curve extracted from DHS substripe~7 on detector NRCA3 from commissioning observations of the WASP-18b transit. Blue points show the individual integrations, while yellow points show the data binned into 10-minute intervals.}
\label{fig:white_lightcurve}
\end{figure}

The resulting dataset consists of simultaneous time-series observations of all DHS spectra on the four short-wavelength detectors together with the long-wavelength grism spectra. The first reduction of the commissioning dataset demonstrates the successful operation of the new observing mode throughout an entire exoplanet transit. White-light curves were extracted independently for all DHS substripes in both the short- and long-wavelength channels. Figure~\ref{fig:white_lightcurve} presents one representative white-light curve obtained from DHS substripe~7 on detector NRCA3 where the transit of WASP-18b is clearly visible. These first on-sky time-series observations demonstrate that the NIRCam DHS multistripe mode is capable of producing stable white time series.

\section{CONCLUSION}
\label{sec:conclusion}
One of the current observational limitations of JWST is the ability to perform high-precision spectroscopy of the brightest targets ($J\sim$3--6 mag), many of which host exoplanets of high interest for atmospheric characterization.
In this paper, we presented the commissioning of a new capability on board JWST, the NIRCam Short Wavelengths Grism Time Series mode. This new mode combines the Dispersed Hartmann Sensor (DHS) optical element with a dedicated multistripe readout. The DHS acts like a set of grating-equipped sub-apertures, that sample approximately 25\% of the primary mirror, substantially reducing the incident flux while simultaneously producing up to 8 short-wavelength spectra of the same target. These spectra fall at different locations on the 4 Module A detectors NRCA1--NRCA4. The implementation of the multistripe readout mode further helps increase the brightness limit. It reduces the frame time from 10.74 s (reading out the full detector) to 1.36 s, by reading out only the detector pixels within given regions of interest containing DHS spectra.
The commissioning campaign successfully validated the detector geometry of the DHS spectra, the capability of the new multistripe readout mode, established wavelength calibration and demonstrated the functionality of the \texttt{jwst} pipeline to process science-ready data. The first time series observations of the transiting exoplanet WASP-18b confirmed the successful operation of the mode.

While the first commissioning milestones have been completed, several activities remain underway, including flux and wavelength calibration for all filter and field-point combinations. On the time-series side, the commissioning analysis is focused on validating the spectroscopic performance of the new mode by measuring the light-curve precision, the uncertainty on the recovered spectroscopic transit depths, and the time-correlated noise across all four detectors and substripes. These analyses require the completion of the end-to-end calibration framework and a detailed assessment of the newly acquired commissioning TSO datasets. Consequently, the present work is intended as a status report on the implementation, calibration strategy, and initial validation of the DHS mode. A comprehensive characterization of the time-series performance, including white- and spectroscopic-light-curve analyses, residual-noise properties, and the construction of final transmission spectra, is currently in progress and will be presented in a dedicated follow-up publication.

More generally, the multistripe readout mode represents a new read out capability that not only applies to NIRCam but to any detector on-board JWST. More than that, it is a very powerful tool to enable the observation of bright targets and avoid saturating the detectors across all instruments while maintaining full observing efficiency. Building on the successful commissioning of this readout mode on NIRCam, similar multistripe implementations are planned for NIRISS/SOSS and NIRSpec/PRISM. As the characterization of bright exoplanetary systems nearby remains a high scientific priority for the JWST community, the development of multistripe readout modes will play an important role in expanding the observatory's capabilities for high-precision spectroscopy.  

\acknowledgments 
This work is based on observations made with the NASA/ESA/CSA James Webb Space Telescope, obtained under Cycle~3 and Cycle-4 programs PID~4453, PID~9215 and PID~9243. The authors thank the Space Telescope Science Institute operations teams for their work and support.

\bibliography{report} 
\bibliographystyle{spiebib} 

\end{document}